# Schrödinger's cat in Einstein's box

Raoul Nakhmanson[1]

*Abstract:*  Using the Einstein's boxes thought experiment, as well as EPR and Heisenberg's ones, the local-realistic hidden-variable interpretation of quantum mechanics is explained. The key hidden variable is the consciousness forecasting the future. It is supposed that atoms and particles are complex products of evolution.

## 1. Introduction.

Thought experiments were very popular in discussions devoted to physical interpretation of quantum mechanics. Some of them, for example the two slits experiment with electrons, "delayed choice", and the Einstein-Podolsky-Rosen (EPR), were later realized in laboratories. In this paper a thought experiment will be discussed which was not and perhaps will never be realized in the laboratories, which will not belittle its heuristic meaning. There is the "Einstein's boxes" (EB) thought experiment. Two recent articles in the February issue of American Journal of Physics demand to reply.

The article [1] "collects together several formulations of this (i.e. EB – *RN*) thought experiment from the literature, analyzes and assesses it from the point of view of the Einstein-Bohr debates, the EPR dilemma, and Bell's theorem, and argues for `Einstein's boxes´ taking its rightful place alongside similar but historically better known quantum mechanical thought experiments such as EPR and Schrödinger's cat" (end of citation).

The article [2] points out that in EB, in the contrary to EPR, "one is not forced to accept non-local causality as the only alternative" and claims "the EPR argument for the incompatibility of locality with the completeness of a quantum-mechanical description of physical reality is definitively stronger than the box (i.e. EB – *RN*) argument."

The articles [1] and [2] pull us back into the first half of the 20th century. Since that time our understanding of physical reality was developing. In particular, there is an opinion that information plays an essential role not only in living beings but also in an "inanimate" matter [3–5]. The presented article discusses the EB thought experiment, as well as EPR and Heisenberg's ones, from this point of view. For history and old discussions of the EB thought experiment see Ref. [1].

## 2. Einstein's boxes thought experiment.

The configuration of EB is shown in Fig. 1.

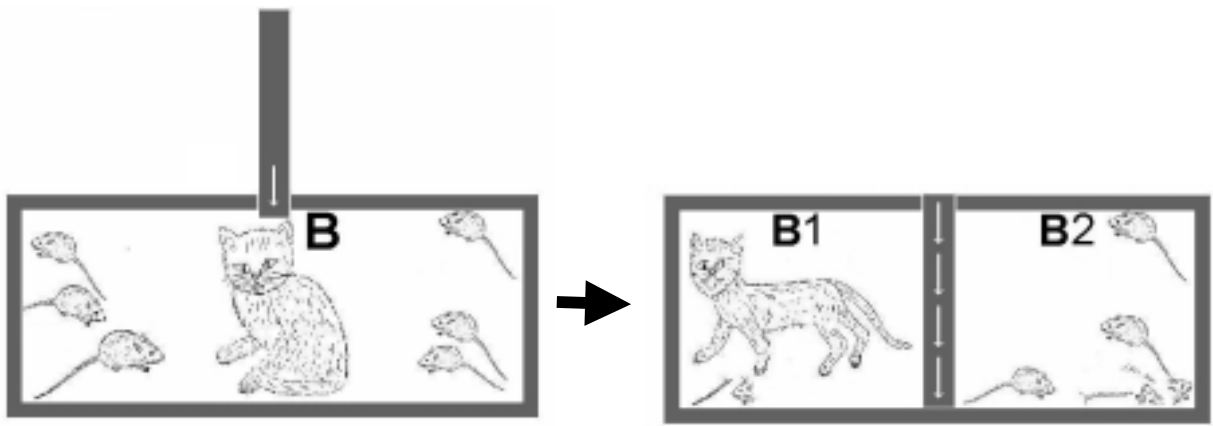

Fig. 1.  Einstein's boxes thought experiment .

---

[1] E-mail address:  nakhmanson@t-online.de

The first step: An object is brought into a big box **B** with impermeable walls. Originally this object was a ball. But objects of quantum physics are atomic i.e. indivisible, that is, it is impossible to split one electron into two smaller electrons with the same topology, etc. The same is valid for living beings. On the contrary the ball made e.g. from iron can be split into two smaller balls, so it is not an appropriate object for us. Therefore I have borrowed from Schrödinger his still alive cat. The wave function of the cat $\Psi(x, y, z, t)$ is confined to the box (idiom), the $|\Psi|^2$ gives a probability to find the cat in the point $(x, y, z)$ at the moment $t$.

The second step: An impermeable sliding shutter penetrates into the box **B** and divides its inner space in two equal parts. Such a procedure divides also the wave function of the cat in two parts (a standard but doubtful conclusion).

The third step: Someone saws the box holding on to the central plane of the sliding shutter. As a result we have two separate small closed boxes **B**1 and **B**2 without looking where the cat is. Of course one must be sure that the cat does not run away.

The fourth step: The box **B**1 stays on the place e.g. in Paris and the box **B**2 is sent off e.g. to Tokyo. The probabilities to find the cat in Paris or in Tokyo are both 1/2.

The fifth and last step: An experimenter opens the box **B**1 and (with probability 1/2) finds the cat. In this case she can immediately conclude that the box **B**2 is empty i.e. the wave function contained in **B**2 immediately collapses to zero in spite of the long distance between **B**1 and **B**2. Apparently we have an instant action over (any) distance that contradicts our physical knowledge and common sense. That is the "Einstein's boxes" paradox. Conclusion: Quantum mechanical description of physical reality is incomplete.

### 3. Heisenberg's thought experiment

Author of [1] remembers that Heisenberg discussed something like EB and quotes from [6]:

". . . one other idealized experiment (due to Einstein) may be considered. We imagine a photon which is represented by a wave packet built up out of Maxwell waves. It will thus have a certain spatial extension and also a certain range of frequency. By reflection at a semitransparent mirror, it is possible to decompose it into two parts, a reflected and a transmitted packet. There is then a definite probability for finding the photon either in one part or in the other part of the divided wave packet. After a sufficient time the two parts will be separated by any distance desired; now if an experiment yields the result that the photon is, say, in the reflected part of the packet, then the probability of finding the photon in the other part of the packet immediately becomes zero. The experiment at the position of the reflected packet thus exerts a kind of action (reduction of the wave packet) at the distant point occupied by the transmitted packet, and one sees that this action is propagated with a velocity greater than that of light."

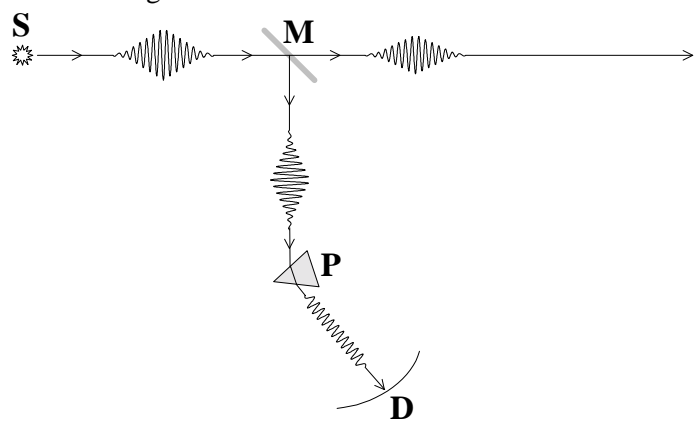

Fig. 2. Heisenberg's thought experiment. **S** is a source of photons,
**M** is a semitransparent mirror, **P** is a prism, **D** is detector.

The scheme of Heisenberg's thought experiment is shown in Fig.2. I have added there in the reflecting alternative a spectrometer as a measurement apparatus. The Heisenberg's photon or wave packet interacts with a prism. As a result of this interaction a certain mode of the latent photon spectrum is chosen and a monochrome photon reaches a corresponding point of a photographic plate or an array of photodiodes. Only now, according to the prevalent Copenhagen interpretation, to which Heisenberg trended, we can speak about a real photon:

> "No elementary phenomenon is a phenomenon until it is a registered (observed) phenomenon. It is wrong to speak of the route of the photon in the experiment of the beam splitter. It is wrong to attribute a tangibility to the photon in all its travel from the point of entry to its last instant of flight." [7]

It is difficult to accept such a solipsism generated by a disability to explain apparent paradoxes. Heisenberg's text seems also inconsistent. His photon "is represented by a wave packet built up out of Maxwell waves". What means "represented"? The wave packet carries energy because it is "built up out of Maxwell waves". After the mirror each of the secondary packets carries half energy, which in quantum mechanics corresponds to a half frequency and contradicts to the experiment. By the way: Not photons built up out of Maxwell waves but, *vice versa*, waves built up out of photons. Also remember Schrödinger's fruitless attempts to build particles from waves. Further, if the photon contained all frequencies of the mother atom's band, it is conclusive to accept it contains all polarizations and all *directions* too. The latter is hard to imagine.

Like Planck, Einstein, de Broglie, and Schrödinger I believe in local realism and incompleteness of quantum-mechanical description. For example the Heisenberg's picture of the photon can be corrected as following: It is a particle having a definitive frequency $\nu$ and energy $h\nu$, its $\nu$-spectrum is a delta-function.

**FAQ:** Delta $\nu$-spectrum corresponds to an infinite space interval. Where is the photon?
**A:** Having the time, point, and direction of the photon after its last interaction you can calculate where and when it is now. This answer is valid for the wave packet too.

A certain range of frequency has not a lonely photon but an ensemble of photons which belongs to a certain atom. The individual frequencies belong to different photons and do not interfere after the photon has given up the mother atom. Only at the moment of birth when the photon with the help of mother atom has a possibility to choose its frequency (and direction, and polarization) there are superposition and interference. After that there is a reduction (collapse) of the superposition state to the delta-function of $\nu$, definitive direction and polarization. The next possibility to have a superposition appears by the interaction with the semitransparent mirror where are reflection/transmission alternatives. After decision (e.g. "reflection" and collapse of "transmission") the photon has no alternatives up to registration by the photographic plate or the array of photodiodes (the reflections on the surfaces of the prism are neglected).

Therefore the interpretation of a single photon as a wave packet is wrong. The "new" (since 1905) interpretation leads to the same result in the experiment Fig. 2 as Heisenberg's one, but is self-consistent and promising.

### 4. Wave function

Now let me analyze the wave function. The first question is "Where is it?" Two extreme cases were presented by physical society:

(1) Wave function (at least its coordinates' part) is in real 3-D space. Such an idea attends in Heisenberg's text, it is presented in EB, and it was tested by de Broglie, Schrödinger, and Bohm.

(2) Wave function is in the consciousness of a physicist thinking about. Such an idea was supported by von Neumann, London and Bauer, and Wigner.

Both cases have their devotees up to now. The disadvantage of the case (1) is non-locality. It is essence of EB, EPR and Heisenberg's examples. In the case (2) it is pointed out that a wave function is not in real but in a configurational space, which is, in its turn, in the consciousness of the physicist i.e. occupying perhaps only a small space part of brain. Therefore the non-locality is not a matter here. The disadvantage of the case (2) is a doubtful connection between object and subject. Were physics the same before we were coming? Sure "Yes". What would change in the object's behavior if the physicist thinks wrong? Sure "Nothing".

Our consciousness receives, works up and spreads information. It controls our behavior to realize our intentions. As it seems our behavior also has a wave component [8]. My plans, my strategy is my wave function. To know a wave function of another person, cat, or electron, I must observe, experimentalize, deduce, guess, and/or ask the object. The original wave function of any object must be in the object itself.

Let us return to the EB thought experiment. At the beginning the cat was mewed into the box **B** together with several mice as food. It surveys the box and works up its wave function. The cat is hungry and seeks for mice. But the mice are not stupid. They guess the cat's wave function and look for a regularity of the cat's trips. If mice find such a regularity they can better get away from the cat. To avoid it the cat randomizes its behavior, that is, if it has alternatives, it chooses randomly with weights recommended by the wave function. Atomism, wave function (strategy), and randomization (tactics) are three whales of quantum mechanics. The peculiarities of the cat's (particle's) behavior lie outside of quantum mechanics. Conclusion: Quantum-mechanical description of physical reality is incomplete. To know these peculiarities we can try to act on atoms and particles with information, try "to speak" with them. The schemes of such experiments were suggested in [3–5].

The second step in EB: An impermeable sliding shutter penetrates into the box **B**. If the shutter hits and kills the cat then its wave function vanishes. The cat tries to evade it. With good luck as an atomic i.e. indivisible object it is in the left *or* in the right half of **B** and later in the box **B**1 *or* in the box **B**2. The idea of a wave function split in half [1] is wrong.

The cat surveys its new box and works up in its consciousness the new wave function. This is the original, right, correct, genuine wave function of the cat. The physicist has not an actual information and cannot imagine the correct copy of original wave function of the cat before **B**1 and/or **B**2 are open.

Therefore there is a very natural explanation of Einstein's boxes thought experiment. We must only accept that the object is indivisible and has a hidden variable like a consciousness. For people, cats, and mice it is obvious. Why not for atoms and particles? As it seems they are very complex products of evolution. Such an acceptance is more natural compared with "*many worlds*", "*empty*" and "*advanced*" waves. Moreover, it includes these ideas as mental alternatives appearing in the object's consciousness. And this consciousness can be tested by experiments [3–5].

### 5. EPR experiment and Bell's theorem

As it is seen from [1, 2], their authors think that nonlocality of quantum interaction is well established by experiments. But Bell's theorem has one logical hole. Namely, it does not take into account such a hypothetical non-mechanical hidden variable which works as a consciousness having a possibility to predict the future. So far as I know such a possibility was firstly point out in [3].

The information available to the particle is the knowledge of the past. For solving the variation problem, the particle must be able to predict what is coming. Prediction is a necessary attribute of any consciousness.

Two or more particles may have a common strategy. In such a case their common wave function does not decompose into a product of partial wave functions. Such "entangled" particles, being separated in space, nevertheless act in a concerted way.

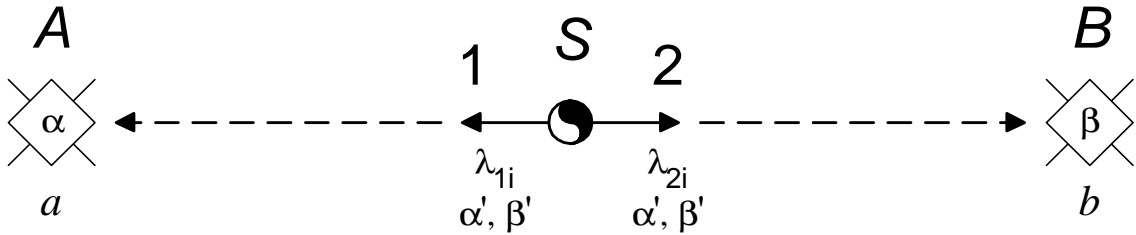

Fig. 3. Einstein-Podolsky-Rosen (EPR) experiment .

Fig. 3 presents the scheme of Einstein-Podolsky-Rosen (EPR) experiment. The source *S* emits two entangled particles **1** and **2** possessing series of hidden parameters $\lambda_{1i}$ and $\lambda_{2i}$, respectively. The particles fly in opposite directions and are measured in distant points *A* and *B*. The measurers have adjusted conditions $\alpha$ and $\beta$, respectively, and give out results *a* and *b*.

The proof of Bell's theorem is based on the next assertion: If $P_a$ is the probability of result *a* measured on the particle **1** in the point *A* having a condition (e.g. angle of analyzer) $\alpha$, and $P_b$ is the probability of result *b* measured on the particle **2** in the distant point *B* having a condition $\beta$, then $\beta$ has no influence on the $P_a$, and *vice versa*. Herein Bell and others saw the indispensable requirement of local realism and "separability". Mathematically it can be written as

$$P_{ab}(\lambda_{1i},\lambda_{2i},\alpha,\beta) = P_a(\lambda_{1i},\alpha) \times P_b(\lambda_{2i},\beta) \quad \text{(Bell)}, \qquad (1)$$

where $P_{ab}$ is the probability of the join result *ab*, and $\lambda_{1i}$ and $\lambda_{2i}$ are hidden parameters of particles **1** and **2** in an arbitrary local-realistic theory. Under the influence of Bell's theorem and the experiments following it and showing, that for entangled particles the condition (1) is no longer valid, some "realists" reject locality. In this case an instantaneous action at a distance is possible, and one can write

$$P_{ab}(\lambda_{1i},\lambda_{2i},\alpha,\beta) = P_a(\lambda_{1i},\alpha,\beta) \times P_b(\lambda_{2i},\beta,\alpha) \quad \text{(nonlocality)}. \qquad (2)$$

In principle such a relation permits a description of any correlation between *a* and *b*, particularly predicted by QM and observed in experiments. But if the particle possesses consciousness then the condition (1) is not indispensable in the frame of local realism. Instead, one can write

$$P_{ab}(\lambda_{1i},\lambda_{2i},\alpha,\beta) = P_a(\lambda_{1i},\alpha,\beta\prime) \times P_b(\lambda_{2i},\beta,\alpha\prime) \quad \text{(prediction)}, \qquad (3)$$

where $\alpha\prime$ and $\beta\prime$ are the conditions of measurements in points *A* and *B*, respectively, as they can be predicted by particles at the moment of their parting. If the prediction is good enough, i.e., $\alpha\prime \approx \alpha$ and $\beta\prime \approx \beta$, then (3) practically coincides with (2) and has all its advantages plus locality. The way for local-realistic description of nature is free.

There are two ways to confirm the consciousness of particles. The first is to prevent the particles from correctly predicting the future, which should lead to nonstandard results. Such attempts were made by groups of Aspect [9], Alley [10], and Zeilinger [11]. The first group used periodically switching of analyzers which was, of course, predictable. Two other groups used the "random" switching, but the randomness was borrowed from the object of study itself (i.e. the quantum world), which cannot be regarded as reasonable. For example, it is not enough to test autocorrelation of quantum random number generator [12], the correlation between two or more such generators is important. Perhaps using a good human-programmed "quasi-random" generator is preferable.

## 6. Informational experiments

The second way is to affect the particles with information. Note that carrying out of such "informational" experiments with elementary particles differs from anything that has been done in physics so far. And, of course, any consciousness can manifest itself only in a situation having alternative outcomes.

It would be safer to assume that we are dealing with a totally different civilization that knows nothing about us, so that our first contacts will run into difficulties. This problem is not new, and has been seriously considered in the framework of the Search for Extraterrestrial Intelligence project SETI. Its experts are inventing "cosmic" languages capable of developing communication from zero to a high semantic level. At the initial stage one could recommend trying such universal languages as mathematics and music. The starting point for identifying intelligence that may be much unlike ours, and for trying to establish contact with it, should be some very general property presumable inherent in any kind of intelligence. A good candidate for such a role is curiosity.

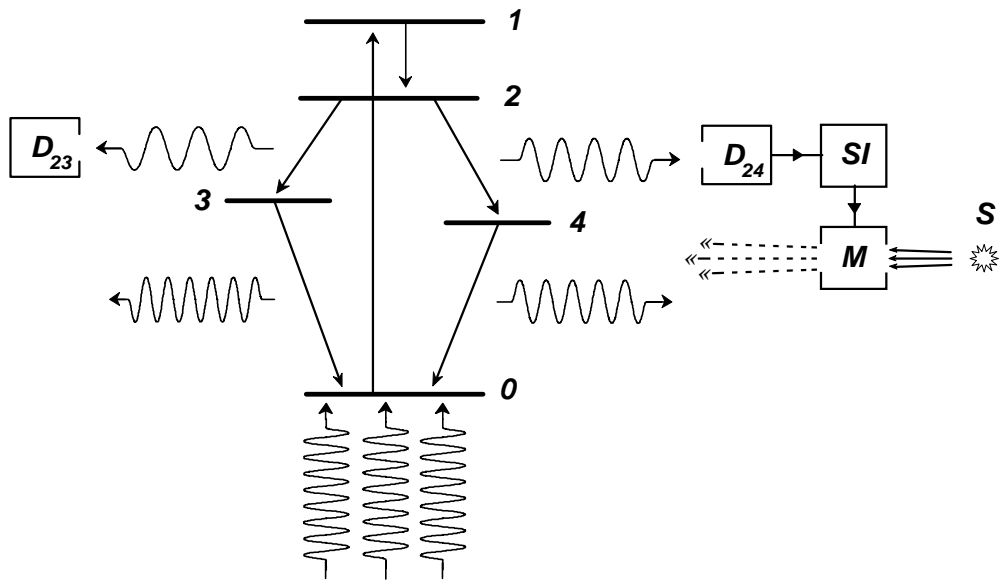

Fig. 4. Informational experiment with a single ion. 0, 1, 2, 3, and 4 are the energy levels; $D_{23}$ and $D_{24}$ are detectors, $S$ is the source of light, $M$ is the modulator, $SI$ is the source of information.

Let me illustrate this with one example. Figure 4 shows the scheme of an experiment that does not presume that the particles are aware of our culture in any way. Here a five-level quantum system, e.g., an individual ion in the trap, with one low (0), one high (1), and three intermediate (2,3,4) energy levels, is pumped by intensive radiation inducing the 0→1 transition, so that the ion does not stay in the state 0 but immediately is translated into the state 1 followed by metastable state 2. From it, the ion makes a spontaneous transition to the states 3 or 4, and later makes a transition to the state 0 completing the cycle. The radiation corresponding to some of the transitions 2→3, 2→4, 3→0, and 4→0 are detected (in Fig. 4 two detectors are shown). Besides, there is an informational action upon the ion, e.g., by modulation of light coming from the source $S$. The modulator $M$ is controlled by the source of information $SI$, which, in turn, is connected with one or more detectors to close the feedback loop.

The feedback works in such a way as to stimulate a channel and rate of transitions, in the case of Fig. 4, the $2\to4$ and $4\to0$ transitions. The source $SI$ sends a message, e.g., one line of a page or a measure of a music, only if it receives a signal from detector $D_{24}$. Each next message continues the previous one, i.e., is the next line or the next measure.

If the ion has a consciousness and is interested in the information being proposed, it develops a conditioned reflex and will prefer the $2\to4$ transition to the $2\to3$ one. Besides, the rates of both $2\to4$ and $4\to0$ transitions must increase. All this can be registered by the experimenter. To be sure that the effect is connected with information, one can make a control experiment to cut off the feedback or/and to use some "trivial" information, etc.

Deviation from the standard transitions may be interpreted as an interest of the ion towards the information. Such an interest is thought to be an inherent attribute of each consciousness. This important result does not even depend on the ability of the ion to decipher information. It is sufficient that it is curious. It is like people of modern times were interested in ancient hieroglyphic symbols long before they learned how to read them.

The sequence of information offered for the ion may be a kind of a course teaching it a language for further dialogue. To measure the progress of learning, the experimenter may sometimes introduce into informational channel some specific "request". For example, one can "ask" the ion to choose $2\to3$ transition rather than the $2\to4$ one. Since the ion, eager not to shirk the lessons, will tend to prefer the $2\to4$ transition, the execution of this request can easily be detected by an experimenter and will mean that the text was decoded.

However, the possibilities of an experiment typified in Fig. 4 are not exhausted by this. Purposefully choosing the transitions, the ion, in its turn, can send information to the experimenter using "right" and "left" (in Fig. 4) transitions as a binary code.

More examples of informational experiments can be found in [3–5]

### 7. Conclusion

The interpretation of quantum mechanics has long history and is under discussion up to now. Two extreme directions are seen. The first is minimalistic and pragmatic. It follows to the principle thought as old as Aristoteles but named after William of Ockham "Ockham's Razor":

"entities should not be multiplied beyond necessity".

The head person of this direction was Niels Bohr. He tried to cancel a very natural, comfortable and loved idea of reality of our world.

But Bohr himself told his students that there are trivial and deep statements. To be asked "What is a deep statement?" he answered: "It is such a statement, that an opposite statement is also a deep one." If one accepts the Ockham's principle as a deep statement then, according to Bohr,

"entities should not be canceled beyond necessity"

must also be accepted as a deep statement. The head person of this second direction was Einstein. He did not cancel reality. His photon born in 1905 is only one example for that.

The "informational" (or "evolutional", or "sociological" [13]) interpretation of quantum mechanics is a bridge between solid grounds of 1905 and 2005 over the desert of solipsism, labyrinths of maths, and jungle of paradoxes. The relativity of subject-object connection accepted in biology seems to be spread on all matter. It opens a way beyond quantum mechanics.